\documentclass[prl, showpacs,preprintnumbers,superscriptaddress,amsmath,amssymb]{revtex4}

\usepackage{graphicx}
\usepackage{color}

\begin{document}

\newcommand{\feq}{\ensuremath{f^{\text{eq}}}}
\newcommand{\bu}{\ensuremath{\mathbf{u}}}
\newcommand{\cc}{\ensuremath{\mathbf{c}}}
\newcommand{\bj}{\ensuremath{\mathbf{j}}}
\newcommand{\bv}{\ensuremath{\mathbf{v}}}
\newcommand{\hmtTwo}{\ensuremath{{ \cal H}^{(2)}}}
\newcommand{\hmtThree}{\ensuremath{{ \cal H}^{(3)}}}
\newcommand{\bvhat}{\ensuremath{\hat{\mathbf{v}}}}
\newcommand{\bx}{\ensuremath{\mathbf{x}}}
\newcommand{\normSt}{\ensuremath{{N}}}
\newcommand{\JCal}{\ensuremath{{{\cal J}}}}
\newcommand{\Kn}{\ensuremath{{\rm Kn}}}
\newcommand{\cs}{c_\text{s}}

\title{ Mean-field  Model beyond Boltzmann-Enskog  Picture for Dense Gases}

\author{Santosh Ansumali}\email{ansumali@gmail.com}
\affiliation{School of Chemical and Biomedical Engineering,
Nanyang Technological University, 639798 Singapore}

\begin{abstract}
{I propose an extension to Boltzmann BGK equation for Hard Spheres. The present model has an $H$-theorem and  it allows  choice of   Prandtl number as an independent parameter.   I show that similar to Enskog equation this equation can reproduce hydrodynamics of hard spheres in dense systems.}
\end{abstract}
\pacs{47.11.-j, 05.70.Ln} \maketitle

\section{Introduction}
An overwhelming majority of fluid flow problems of physical and engineering interest cannot be solved using microscopic simulation methods, such as molecular dynamics, due to the enormous number of degrees of freedom constituting the macroscopic systems.  In such a scenario, mesoscale  descriptions in terms of one particle distribution function, such as Boltzmann equation,  provide   important tools for understanding transport phenomena beyond   phenomenological hydrodynamic descriptions in terms of Navier-Stokes-Fourier equations.    Indeed,  the nonlinear Boltzmann kinetic equation can accurately predict  a wide range of physical properties and flow profiles for low density gases even in states   very far from equilibrium (see for example \cite{cercignani1988bea}). 

However,    technical difficulties encountered in solving (analytically or numerically) the Boltzmann equation, a nonlinear integro-differential equation for the time dependent distributions in a six dimensional phase space, limits its application in practice.  During the last few decades, this technical problem has been solved for the Boltzmann equation in two very important regimes. Firstly, for highly non-equilibrium situations associated with supersonic flows (in general for high Mach number flows), direct simulation  Monte Carlo (DSMC) method  was applied with remarkable success (see for reviews\cite{bird1994mgd,oran1998dsm}). Secondly, for very low Mach number flows  lattice Boltzmann method is remarkably successful in both hydrodynamic regime as well as transitional regime (see for example \cite{chen2003ebk,succi2001lbe, ansumali2007hbn}). The lattice Boltzmann method rely on an approximate form of Boltzmann collision term known as Bhatanager-Gross-Krook (BGK) collision approximation. The model Boltzmann equation with BGK collision 
term retains almost all qualitative features (such as correct conservation laws, $H$-theorem)   of the Boltzmann equation \cite{bhatnager1954mcp}.  Indeed, the BGK model can be classified as the first truely successful phenomenological model at the level of one particle distribution. The mathematical simplicity of this model is often   used to obtain exact and semi-exact analytical solutions which can help to understand the hydrodynamics well beyond Navier-Stokes equations \cite{cercignani1988bea, ansumali2007hbn}.  The strength and limitations of this model   along with ways to make  it  quantitatively accurate (without destroying the basic features such as $H$-theorem) is  well understood  \cite{cercignani1988bea,gorban1994gac,gorban2005imp,ansumali2007qel}. 
 
For hard spheres, in order to describe the  fluid transport in dense regimes,   Boltzmann equation    was extended by Enskog  and further modified   by van Beijeren and Ernst  (known as the revised Enskog theory (RET)) \cite{chapman1991mtn,vanbeijeren1973mee}. 
  Similar to Boltzmann's model of diluete gas,  particles motion in these models is decomposed into two parts: propogation at constant velocity followed by  collisions in which   exchange of momentum between particles happens. However, unlike Boltzmann model, collisions are understood to be  non-local events due to the presence of  finite size particles.  This idea of non-local collisions due to finite size of particles  behind Enskog or RET extension of Boltzmann equation is borrowed from Van-der Waals' picture of excluded volume in dense gases. However, mainly  due to the  non-local collisions,  Enskog extension of  Boltzmann model leads to even more  intractable form of  nonlinear integro-differential equation.  Thus, it is not surprising that it took almost fifty  years to prove even the existence of $H$-theorem \cite{resibois1978htm} and so far only modest engineering and physical application of  dense fluid is modeled  via Enskog equation. The non-local nature of collision is difficult to handle both for  Monte-Carlo method as well as for kinetic modeling via simplified phenomenological theories of  BGK type.

From engineering perspective, a dense gas model, where non-ideality can be  added as extra terms over a rarefied gas model (either of Monte-Carlo type or BGK type), is an extremely desirable  solution. In case of  DSMC model of Boltzmann equation, Alexander et. al proposed a simple modification of propagation step  which gave the correct
equation of state but failed to reproduce the Enskog
transport coefficients  \cite{alexander1995cba}. An important progress for such modeling approaches  was reported in Ref.\cite{dufty1996pkm,lutsko1997ase}, where   BGK--like collision terms for Enskog equation were proposed.  It needs to be remarked that  most   multiphase extensions of lattice Boltzmann method uses  simplifications pertinent to incompressible flow in  Dufty et. al model \cite{ dufty1996pkm} to  mimic hard-sphere repulsion (see \cite{succi2001lbe} for details of multiphase lattice Boltzmann method). 
The main idea  behind these works   was to compute the effect of Enskog collision term on momentum and energy balance and explicitly add it in momentum and energy balance equation as a correction to the BGK collision term. This approach gave correct viscosity  coefficient and equation of state  but  there is neither  $H$ theorem nor correct thermal conductivity for these type of  models.  

The goal of the present work is to fill this gap and construct a phenomenological model of fluid transport at the meso-scale level. Similar to hydrodynamic description given by Navier-Stokes equations, I demand that a good phenomenological theory at mesoscale should fullfil the following criteria
\begin{itemize}
\item{It must obey conservation laws and second law of thermodynamics($H$-theorem).}
\item{It must reproduce correct thermodynamic equation of state.}
\item{It must reproduce correct Enskog transport coefficients in hydrodynamic regime.}
\end{itemize}

   The basic starting point for such  modeling is   exact dynamics of one particle distribution function $f$ (defined as probability of finding a particle  at location $\bx$ with velocity $\bv$ at time $t$) as  a function of phase variable ${\mathbf z}\equiv (\bx, \bv)$,  given by BBGKY hierarchy as
   \begin{equation}
   \partial_t \,f\left({\mathbf z}_1 , t\right) +  \partial_\alpha\left [f\left({\mathbf z}_1, t\right)  \, \bv_1 \right] =-\hat{A}_1\, f_2({\mathbf z}_1,{\mathbf z}_2, t)   \end{equation}
   where $f_2(1,2)$ denotes the two particle distribution function and $\hat{A}_1$ is an operator whose exact form is not important for the present discussion (For details of this equation see for example \cite{liboff2003ktc}).   and the 
hydrodynamic fields  $\left\{\rho, \bj,T\right\}$ are defined in terms of one particle distribution function $f$ as
\begin{equation}
\int d\bv\, f\, \left\{1,\bv, \frac{v^2}{2}\right\}=\left\{\rho, \bj, \frac{j^2}{2\,\rho}+ \frac{\rho \, D\,R\,T}{2}\right\}.
\end{equation}
   A closed form kinetic equation is obtained if I set $\hat{A}_1\, f_2=\JCal(f)$, where $\JCal$, typically collision integral, maps functions onto functions. Thus for example, $f_2= f_2(1,2 , f)$ with explicit time dependence of $f_2$ entirely contained in $f$ was proposed by Bogoliubov to derive the Boltzmann equation. Traditional mechanistic view-point starting from  BBGKY hierarchy is that non-ideality coming from repulsive part of molecular forces needs to be modeled via    three-particle and other higher order collision and free-flight of molecules remains unaffected by it.  However, so far only widely used model is mean-field approximation  of Enskog, where all higher order effects are lumped in non-local collisions.  In the present work, I  am advocating an alternate mean-field  picture of non-ideality in dense system. According to this picture in dense system, dominant change is alteration of free flight of a hard sphere particle.
 The key new ingradient, I propose  is that $\hat{A}_1\, f_2$ will also modify the free propogation step of the Boltzmann type equations. In particular, I propose to write generalised kinetic equation for  $f({\mathbf z},t)$ as 
   \begin{equation}
\label{timeevolution}
\partial_t \,f\left({\mathbf z}, t\right) +  \partial_\alpha\left [f\left({\mathbf z}, t\right)  \, \bvhat\right] 
 =  \JCal 
\end{equation}
where $\bvhat$ is some unknown propagation velocity.   The physical picture behind such a modification of free propagation can be understood using  Bogoliubov hypothesis for the equilibration of a non-equilibrium gas, which assumes separations 
 of time scale by considering mean free time $\tau$ of a molecule to be much larger than mean time spent in the  interaction domain of the another molecule \cite{liboff2003ktc}. In dilute region, where  reduced density is small, using this hypothesis he derived Boltzmann equation. Now for hard-sphere in dense region the dynamics in mean free time $\tau$  has to be more complex due to collective effects. In this time scale, for dense system we also need to consider the effect of collective motion (hydrodynamics) on individual particles.    Such corrections from local collision can also be incorporated by analyzing the short time motion of tagged particle  due to entropic force generated by particle-particle correlations and  hydrodynamics force  generated by other particles.  In other words, apart from usual hydrodynamic forces  the tagged particle  might also experience   effective  forces, which   have  purely entropic origin.  Unlike ring kinetic theory approach, I am not trying to derive such an equation from the first principle but trying to model it based on physical intuitions and limiting behaviors of the system.  Firstly,  we know the two  limiting behaviors of such a system in rarefied and extremely dense regime quantified by compressibility factor $\chi$ defined as
\begin{equation}
  \chi =\frac{p}{\rho \, R\,T}-1\equiv  \frac{1}{\rho \, R}\left(s^{\rm nid} -\rho \frac{\partial s^{\rm nid}}{\partial \rho}\right),
  \end{equation}
  where the excess entropy $s^{\rm nid}\left(\rho\right)$ and pressure $p(\rho,T)$ as a function of density $\rho$ and temperature $T$  is known from equilibrium statistical mechanics (For example Van-der-Waals or Carnhann-Starling approximation). 
  In a rarefied system $(\chi\rightarrow 0)$, present model  should recover Boltzmann description of free propagation step ($\bvhat=\bv$), while for extremely dense system $(\chi\rightarrow \infty)$, all hard-spheres should pack together and move with the collective velocity of the system, which means $\bvhat=\bv=\bu$.

 In the present work, I ask the question what is the most general form of  the propagation velocity $\bvhat$ (which can be a function of the moments of $f$)  in Eq. \eqref{timeevolution}. 
   In order to do so, first  we need to recognize that the propagation velocity can be written as a formal Hermite expanion  $\bvhat  = \bv+ \sum_{ n=0}^{\infty}{\mathbf a}^{(n)} \,  {\cal {\mathbf H}}^{(n)}( {\pmb \xi})$, in terms of  dimensionless peculiar velocity   
$\xi_{\alpha}= \left(v_{\alpha} -u_{\alpha}\right)/\sqrt{2 \,R \,T} $. Furthermore,  we  need to recognize that  the condition of having correct  continuity equation (obtained by integrating Eq.\eqref{timeevolution} over $\bv$) itself severely restricts the choice of $\bvhat$.
For example, ${\mathbf a}^{(0)}=0$ and ${\mathbf a}^{(n)}=0$ for all $n>2$.  For $n=2$ only the trace part  survives and  the most general form of $\bvhat$, consistent with the conservation laws, is 
\begin{equation}
 \label{mainEq}
\hat{v}_{\alpha} -v_{\alpha} = \underbrace{\chi\, \left(v_{\alpha} -u_{\alpha}\right) }_{A} +
 \underbrace{\left(v_{\beta} -u_{\beta}\right)  \, \frac{P^{\rm (I)}_{\alpha \beta}}{\rho \, R \, T} }_{B}
+ \underbrace{u^{\rm (I)}_{\alpha}\,   \left(\xi^2-\frac{D }{2}\right)}_{C}, 
 \end{equation}
  where   precise form of second order tensor  $P^{\rm (I)}_{\alpha \beta}$ and fictitious velocity $u^{\rm (I)}_{\alpha}$ need to be determined  with the requirements of correct   $H$-Theorem. 
  In subsequent section, I will show that these conditions  are fulfilled if 
  \begin{equation}
\label{LIT}
u_{\alpha}^{\rm (I)}= -\lambda^{(\rm q)} \, \tau\, T\, \partial_{\alpha} \, \log{T},\qquad 
 P^{\rm (I)}_{\alpha \beta}= -k_1\, \tau\, \left(\partial_{\beta}\, u_{\alpha}+\partial_{\alpha} \, u_{\beta} -\frac{2}{D}\partial_{\gamma} u_{\gamma} \delta_{\alpha\beta} \right)-k_2\, \tau\, \partial_{\gamma} u_{\gamma} \delta_{\alpha\beta},
\end{equation}
where $k_1$, $k_2$ and $\lambda^{(\rm q)}$ are positive definite scalars related to transport coefficients of the fluid.
  Here, it needs to be noted that the physical picture behind the present model envisions that both  $P^{\rm (I)}_{\alpha \beta}$ and $u_{\alpha}^{\rm (I)}$ are corrections of the order of mean free time $\tau$ .  We can relate these new transport coefficients with that of real fluid by performing the Chapman-Enskog expansion.   The kinetic equation  Eq.\eqref{timeevolution} along with expression of $\bvhat$  given by Eqs.\eqref{mainEq} and  \eqref{LIT},  are the main result of this work.  
  
  It is easy to check that the present formulation gives correct conservation laws for dense gas. This can be seen  by taking moments of the kinetic equation Eq.\eqref{timeevolution}, which yields time evolution equation for the locally conserved fields as
\begin{equation}
\begin{split}
\label{complete1}
\partial_t \, \rho + \partial_\alpha \,j_{\alpha}&= 0, \\
\partial_t \, j_{\gamma}+ \partial_{\alpha} \, \left[
\rho \, u_{\alpha} \, u_{\gamma} +p\, \delta_{\alpha \gamma} 
+\sigma_{\alpha \gamma} 
\right]&=0 \\
 \partial_t \, \left[
\frac{\rho}{2} \, u^2 +e
\right]+\partial_{\alpha}\, \left[\left(\frac{\rho}{2} \, u^2 +e+p\right) \, u_{\alpha} +  
  \sigma_{\alpha \gamma} \, u_{\gamma}  +
q_{\alpha} 
\right] &=0
 \end{split}
\end{equation}
where  
\begin{equation}
\sigma_{\alpha \gamma} =\left(1+ \chi\right)\, \sigma_{\alpha \gamma}^{(\rm K)} +
  P^{\rm (I)}_{\alpha \gamma}+  P^{\rm (I)}_{\alpha \beta}\,\frac{\sigma^{(\rm K)}_{\beta \gamma}}{\rho\,R\,T} +  u_{\alpha}^{\rm (I)}\,\left( \frac{q_{\gamma}^{(\rm K)}}{R\,T}  \right),
\end{equation}
\begin{equation}
q_{\alpha} =\left \{ \left(1+\chi\right) \,
\delta_{\alpha \beta}+ 
\frac{P^{(I)}_{\alpha \beta}}{\rho \, R \, T} \right\}\,  
q_{\beta}^{(\rm K)}   +e \, u_{\alpha}^{\rm (I)}+R\,T\,u_{\alpha}^{\rm (I)}\, \int  d \bv  \, f\left[ \xi^2\,\left(  \xi^2 -\frac{D+2}{2}\right)\right]. 
\end{equation}
The kinetic part of the stress tensor and heat flux are conveniently defined in terms of
  second and third order traceless Hermite tensor 
\begin{equation}
\hmtThree_{\alpha}= \xi_{\alpha} \,\left( \xi^2-\frac{D+2}{2}  \right),\qquad 
\hmtTwo_{\alpha \beta}=\left(\xi_\beta  \, \xi_{\alpha}-  \frac{1}{ D}\, \xi^2 \, \delta_{\alpha\beta} \right)
\end{equation}
 In terms of these tensors kinetic part of the stress tensor is 
 $
\sigma_{\alpha \beta}^{(\rm K)}=  2\, R\, T\, \int  d\bv f \, \hmtTwo_{\alpha \beta}
$
and kinetic part of the heat flux is 
$
q_{\alpha}^{(\rm K)}  
= {\left(2\, R\, T\,\right)^{3/2}}\, \int  d\bv f /{2}\, \hmtThree_{\alpha}
$. We can show that    the  stress tensor and heat flux has same expression as RET if $k_1$,  $k_2$ and $\lambda^{(\rm q)}$ are chosen properly.   By performing Chapman-Enskog expansion, equation \eqref{timeevolution} provides  the first correction from Maxwell-Boltzmann distribution  as 
  
 
 
  
   \begin{equation}
{\cal L} f^{(1)} =f^{\rm eq}\, \left[
 2\, (1+\chi)\, \hmtTwo_{\alpha \beta} \, \partial_\alpha \, u_{\beta} +
\sqrt{2\, R\, T}\, (1+\chi)\,\hmtThree_{\alpha}
  \partial_{\alpha} \,\log{T}  \right],
\end{equation}
where ${\cal L}$ is the linearized collision operator.  At this stage, we need to provide the exact form of collision operators.  For example, by using BGK collision operator  we obtain

\begin{equation}
\sigma_{\alpha \beta} =  -
  \eta\,  \left(\partial_{\beta}\, u_{\alpha}+\partial_{\alpha} \, u_{\beta} -\frac{2}{D}\partial_{\gamma} u_{\gamma} \delta_{\alpha\beta} \right)-  k_2\, \tau\, \partial_{\gamma} u_{\gamma} \delta_{\alpha\beta} ,
\end{equation}
 where shear viscosity is 
$
\eta=  \left( k_1+ \left(1+\chi\right) \, p\right)\, \tau  $ and $k_2\, \tau$ is the bulk viscosity.   Similarly for heat flux
 \begin{equation}
q_{\alpha} =  -   \tau  \left\{\frac{D}{2}\, \rho\,T\,\lambda^{(\rm q)}+ p\, (1+\chi)\, C_{p}\right\} \partial_{\alpha} \, {T}
\end{equation}
where specific heat at constant pressure $C_p= (D+2)/2\, R$.    For BGK model, we have more tuning  parameters related to transport coefficient then needed. So, without any loss of generality, we  can set $k_1=0$.
So,
Prandtl number is
\begin{equation}
1-{\rm Pr} = \frac{D}{2\, p\, C_{p}\, \left(1+\chi\right)}\, \rho\,T\,\lambda^{(\rm q)} {\rm Pr}
\end{equation}
  This means  unlike Dufty et al \cite{dufty1996pkm}, in present model we are allowed to set heat conductivity and viscosity coefficients independently.  Thus, if it can be shown that the present model admits $H$-theorem, it can be claimed that this model is first complete phenomenological model for describing dense gas hydrodynamics at the mesoscale level.
  
  Before proving  $H$-theorem, let us try to understand the physical meaning of Eq.\eqref{mainEq}. 
  The physical picture behind first term   on the right hand side of Eq.\eqref{mainEq}  (underlined term $A$) can be understood   in the framework of  social force model created by Helbing and Moln\'ar in the context of traffic dynamics \cite{helbing1995sfm}.      The goal of this term is to keep tagged particle separated  from other particles, and such crossing of particles can be avoided (only in an average sense)   by introducing a force which tries to remove them from dense regions.  What we want is that  at the   time $\tau$ after any collision, the probability of having a particle in denser region is smaller.   A system mimicking such a motion is trajectory of  an individual moving on a street trying to avoid being very close to crowded region after time $t$.   Suppose he is moving very fast in the region of high density,  then  he can avoid  being in the region of high density  at the end of time $t$  by moving   even more faster. In other words, he experiences a social force which accelerates him  in the region of high density if he is moving too fast. In the opposite limit, where the individual is moving too slow compared to the crowd in dense regions. He can avoid being in the region of high density  at the end of time $t$, if he get decelerated  in the regions of high density.   For hard-spheres such an effect can be modeled by underlined term $A$ in Eq.\eqref{mainEq}.  The third term $C$ is an hydrodynamic effect and can be understood as microscopic analog of thermophoretic forces \cite{goldhirsch1983ttg}.  The difference from 
  macroscopic expression is that in stead of transport coefficient a direct dependence on heat velocity appears. This may be due to the fact that the current description (Eq.\eqref{mainEq}) is for short-time motion, whereas transport coefficients appear only in long time limit.   Similarly, the term $B$ in Eq.\eqref{mainEq} is an hydrodynamic force which reflect the tendency to resist locally generated flow field (shear stresses and compression).  

    In order to show the existence of $H$-theorem for the present model, I define  the $H$-function for dense system as
  \begin{equation}
H  \left(\bx,  t\right)= \int d\bv f\left(\bx, \bv,t\right)\, \left[ \log{f\left(\bx, \bv,t\right)}-1 \right]-
\frac{s^{\rm nid}\left(\rho  \left(\bx,  t\right) \right) }{R}.
\end{equation}
This choice is   similar to that   used in proving $H$-theorem for Enskog equation and is motivated from the work of Gremla et al \cite{grmela1980cek}. 
Multiplying Eq.\eqref{timeevolution} by $\log{f}$ and noting that non-ideal part of entropy is a function of density only, we obtain time evolution equation for the  $H$ function as

 \begin{equation}
    \label{HThm}
  \partial_t\,  H +J_H       =  \int \JCal  \log{f} d\bv +\frac{D}{2}\rho \, u_{\alpha}^{\rm (I)}\, \partial_{\alpha} \, \log{T}+
\left(  \frac{P^{\rm (I)}_{\alpha \beta}}{  R \, T} \right)\, \partial_{\alpha} u_{\beta}  
  \end{equation}

where the flux of $H$-function (analog of entropy flux)  is 
\begin{equation}
J_H  =   -\partial_{\alpha} \left(\frac{s^{\rm nid}}{R} u_{\alpha}\right)+ \partial_\alpha\int d\bv\left [f(\log{f}-1)    \, \hat{v}_{\alpha} \right]
\end{equation}
It is interesting to see here that  similar to $H$-function  the flux of it also has a contribution totally dependent  on macroscopic variable $\rho$ and $\bj$. 
For $H$-theorem to be valid,  entropy production should be positive, which means 
right hand side of Eq.\eqref{HThm} must be negative. For Boltzmann collision term or BGK collision term, the first term on the right  term is negative. However, this is possible for the last two terms  if 
Eq. \eqref{LIT} is valid.  Thus, we have proved the $H$-theorem. Here, it is interesting to note that the as compared to Boltzmann kinetic theory, the  new ingredient  required to prove $H$-theorem is just the same as that used in linear irreversible thermodynamics (see for a modern perspective \cite{ottinger2005bet}).

To conclude, in the present manuscript  I have presented an alternate mean-field  model of hydrodynamics at the mesoscale level.    
 Finally, a further interesting simplification  of the model happens if we set $k_1=k_2=\lambda^{(\rm q)}=0$. In this limit, using Eq. \eqref{mainEq}  we have  
   \begin{equation}
   \label{LBMult}
\partial_t \,f  +  v_{\alpha}\, \partial_\alpha f  
 = \frac{1}{\tau}\left[ f^{\rm eq}\left (\rho, \bj, T\right) -f\right]- 
\underbrace{\partial_\alpha\left [f    \left(v_{\alpha} -u_{\alpha}\right)  \,\chi \right]}_{G} 
\end{equation}
This   simplified model  can be suitable for numerical implementations  in lattice Boltzmann form. In particular, if  the distribution function $f$ appearing in the term $G$ on the right hand side of  Eq.\eqref{LBMult} is replaced by $f^{\rm eq}$, it resembles current implementation of hard sphere dynamics in lattice Boltzmann method. Thus, present model provide a self consistent framework for hard sphere dynamics in lattice Boltzmann method. Furthermore, similar model can be formulated with Boltzmann collision form, which will be very useful for  DSMC simulations. 
\acknowledgements{I am dedicating this work to Dr. I. V. Karlin, teacher, well wisher and friend, in fond memory of the unorthodox but comprehensive education in kinetic theory he gave me. I want to thank Professors H. C. \"Ottinger and Dr. V. Kumaran, also my former teachers, for the solid training in non-equilibrium statistical physics and fluid mechanics they imparted to me. I am also thankful to Dr. Rochish Thaokar and Dr.   S. K. Kwak    for the insightful reading of this draft. }
 \bibliography{MultPhaseEnskog}

\begin{thebibliography}{21}
\expandafter\ifx\csname natexlab\endcsname\relax\def\natexlab#1{#1}\fi
\expandafter\ifx\csname bibnamefont\endcsname\relax
  \def\bibnamefont#1{#1}\fi
\expandafter\ifx\csname bibfnamefont\endcsname\relax
  \def\bibfnamefont#1{#1}\fi
\expandafter\ifx\csname citenamefont\endcsname\relax
  \def\citenamefont#1{#1}\fi
\expandafter\ifx\csname url\endcsname\relax
  \def\url#1{\texttt{#1}}\fi
\expandafter\ifx\csname urlprefix\endcsname\relax\def\urlprefix{URL }\fi
\providecommand{\bibinfo}[2]{#2}
\providecommand{\eprint}[2][]{\url{#2}}

\bibitem[{\citenamefont{Cercignani}(1988)}]{cercignani1988bea}
\bibinfo{author}{\bibfnamefont{C.}~\bibnamefont{Cercignani}},
  \emph{\bibinfo{title}{{The Boltzmann Equation and Its Applications}}}
  (\bibinfo{publisher}{Springer}, \bibinfo{year}{1988}).

\bibitem[{\citenamefont{Bird}(1994)}]{bird1994mgd}
\bibinfo{author}{\bibfnamefont{G.}~\bibnamefont{Bird}},
  \emph{\bibinfo{title}{{Molecular gas dynamics and the direct simulation of
  gas flows}}} (\bibinfo{publisher}{Oxford University Press New York},
  \bibinfo{year}{1994}).

\bibitem[{\citenamefont{Oran et~al.}(1998)\citenamefont{Oran, Oh, and
  Cybyk}}]{oran1998dsm}
\bibinfo{author}{\bibfnamefont{E.}~\bibnamefont{Oran}},
  \bibinfo{author}{\bibfnamefont{C.}~\bibnamefont{Oh}}, \bibnamefont{and}
  \bibinfo{author}{\bibfnamefont{B.}~\bibnamefont{Cybyk}},
  \bibinfo{journal}{Annual Reviews in Fluid Mechanics}
  \textbf{\bibinfo{volume}{30}}, \bibinfo{pages}{403} (\bibinfo{year}{1998}).

\bibitem[{\citenamefont{Chen et~al.}(2003)\citenamefont{Chen, Kandasamy,
  Orszag, Shock, Succi, and Yakhot}}]{chen2003ebk}
\bibinfo{author}{\bibfnamefont{H.}~\bibnamefont{Chen}},
  \bibinfo{author}{\bibfnamefont{S.}~\bibnamefont{Kandasamy}},
  \bibinfo{author}{\bibfnamefont{S.}~\bibnamefont{Orszag}},
  \bibinfo{author}{\bibfnamefont{R.}~\bibnamefont{Shock}},
  \bibinfo{author}{\bibfnamefont{S.}~\bibnamefont{Succi}}, \bibnamefont{and}
  \bibinfo{author}{\bibfnamefont{V.}~\bibnamefont{Yakhot}},
  \bibinfo{journal}{Science} \textbf{\bibinfo{volume}{301}},
  \bibinfo{pages}{633} (\bibinfo{year}{2003}).

\bibitem[{\citenamefont{Succi}(2001)}]{succi2001lbe}
\bibinfo{author}{\bibfnamefont{S.}~\bibnamefont{Succi}},
  \emph{\bibinfo{title}{{The Lattice Boltzmann Equation for Fluid Dynamics and
  Beyond}}} (\bibinfo{publisher}{Oxford University Press},
  \bibinfo{year}{2001}).

\bibitem[{\citenamefont{Ansumali
  et~al.}(2007{\natexlab{a}})\citenamefont{Ansumali, Karlin, Arcidiacono,
  Abbas, and Prasianakis}}]{ansumali2007hbn}
\bibinfo{author}{\bibfnamefont{S.}~\bibnamefont{Ansumali}},
  \bibinfo{author}{\bibfnamefont{I.}~\bibnamefont{Karlin}},
  \bibinfo{author}{\bibfnamefont{S.}~\bibnamefont{Arcidiacono}},
  \bibinfo{author}{\bibfnamefont{A.}~\bibnamefont{Abbas}}, \bibnamefont{and}
  \bibinfo{author}{\bibfnamefont{N.}~\bibnamefont{Prasianakis}},
  \bibinfo{journal}{Physical Review Letters} \textbf{\bibinfo{volume}{98}},
  \bibinfo{pages}{124502} (\bibinfo{year}{2007}{\natexlab{a}}).

\bibitem[{\citenamefont{Bhatnager et~al.}(1954)\citenamefont{Bhatnager, Gross,
  and Krook}}]{bhatnager1954mcp}
\bibinfo{author}{\bibfnamefont{P.}~\bibnamefont{Bhatnager}},
  \bibinfo{author}{\bibfnamefont{E.}~\bibnamefont{Gross}}, \bibnamefont{and}
  \bibinfo{author}{\bibfnamefont{M.}~\bibnamefont{Krook}},
  \bibinfo{journal}{Phys. Rev} \textbf{\bibinfo{volume}{94}},
  \bibinfo{pages}{511} (\bibinfo{year}{1954}).

\bibitem[{\citenamefont{Gorban and Karlin}(1994)}]{gorban1994gac}
\bibinfo{author}{\bibfnamefont{A.}~\bibnamefont{Gorban}} \bibnamefont{and}
  \bibinfo{author}{\bibfnamefont{I.}~\bibnamefont{Karlin}},
  \bibinfo{journal}{Physica A} \textbf{\bibinfo{volume}{206}},
  \bibinfo{pages}{401} (\bibinfo{year}{1994}).

\bibitem[{\citenamefont{Gorban and Karlin}(2005)}]{gorban2005imp}
\bibinfo{author}{\bibfnamefont{A.}~\bibnamefont{Gorban}} \bibnamefont{and}
  \bibinfo{author}{\bibfnamefont{I.}~\bibnamefont{Karlin}},
  \emph{\bibinfo{title}{{Invariant Manifolds For Physical And Chemical
  Kinetics}}} (\bibinfo{publisher}{Springer}, \bibinfo{year}{2005}).

\bibitem[{\citenamefont{Ansumali
  et~al.}(2007{\natexlab{b}})\citenamefont{Ansumali, Arcidiacono, Chikatamarla,
  Prasianakis, Gorban, and Karlin}}]{ansumali2007qel}
\bibinfo{author}{\bibfnamefont{S.}~\bibnamefont{Ansumali}},
  \bibinfo{author}{\bibfnamefont{S.}~\bibnamefont{Arcidiacono}},
  \bibinfo{author}{\bibfnamefont{S.}~\bibnamefont{Chikatamarla}},
  \bibinfo{author}{\bibfnamefont{N.}~\bibnamefont{Prasianakis}},
  \bibinfo{author}{\bibfnamefont{A.}~\bibnamefont{Gorban}}, \bibnamefont{and}
  \bibinfo{author}{\bibfnamefont{I.}~\bibnamefont{Karlin}},
  \bibinfo{journal}{The European Physical Journal B-Condensed Matter and
  Complex Systems} \textbf{\bibinfo{volume}{56}}, \bibinfo{pages}{135}
  (\bibinfo{year}{2007}{\natexlab{b}}).

\bibitem[{\citenamefont{Chapman and Cowling}(1991)}]{chapman1991mtn}
\bibinfo{author}{\bibfnamefont{S.}~\bibnamefont{Chapman}} \bibnamefont{and}
  \bibinfo{author}{\bibfnamefont{T.}~\bibnamefont{Cowling}},
  \emph{\bibinfo{title}{{The Mathematical Theory of Non-uniform Gases: An
  Account of the Kinetic Theory of Viscosity, Thermal Conduction and Diffusion
  in Gases}}} (\bibinfo{publisher}{Cambridge University Press},
  \bibinfo{year}{1991}).

\bibitem[{\citenamefont{van Beijeren and Ernst}(1973)}]{vanbeijeren1973mee}
\bibinfo{author}{\bibfnamefont{H.}~\bibnamefont{van Beijeren}}
  \bibnamefont{and} \bibinfo{author}{\bibfnamefont{M.}~\bibnamefont{Ernst}},
  \bibinfo{journal}{Physica} \textbf{\bibinfo{volume}{68}},
  \bibinfo{pages}{437} (\bibinfo{year}{1973}).

\bibitem[{\citenamefont{R{\'e}sibois}(1978)}]{resibois1978htm}
\bibinfo{author}{\bibfnamefont{P.}~\bibnamefont{R{\'e}sibois}},
  \bibinfo{journal}{Physical Review Letters} \textbf{\bibinfo{volume}{40}},
  \bibinfo{pages}{1409} (\bibinfo{year}{1978}).

\bibitem[{\citenamefont{Alexander et~al.}(1995)\citenamefont{Alexander, Garcia,
  and Alder}}]{alexander1995cba}
\bibinfo{author}{\bibfnamefont{F.}~\bibnamefont{Alexander}},
  \bibinfo{author}{\bibfnamefont{A.}~\bibnamefont{Garcia}}, \bibnamefont{and}
  \bibinfo{author}{\bibfnamefont{B.}~\bibnamefont{Alder}},
  \bibinfo{journal}{Physical Review Letters} \textbf{\bibinfo{volume}{74}},
  \bibinfo{pages}{5212} (\bibinfo{year}{1995}).

\bibitem[{\citenamefont{Dufty et~al.}(1996)\citenamefont{Dufty, Santos, and
  Brey}}]{dufty1996pkm}
\bibinfo{author}{\bibfnamefont{J.}~\bibnamefont{Dufty}},
  \bibinfo{author}{\bibfnamefont{A.}~\bibnamefont{Santos}}, \bibnamefont{and}
  \bibinfo{author}{\bibfnamefont{J.}~\bibnamefont{Brey}},
  \bibinfo{journal}{Physical Review Letters} \textbf{\bibinfo{volume}{77}},
  \bibinfo{pages}{1270} (\bibinfo{year}{1996}).

\bibitem[{\citenamefont{Lutsko}(1997)}]{lutsko1997ase}
\bibinfo{author}{\bibfnamefont{J.}~\bibnamefont{Lutsko}},
  \bibinfo{journal}{Physical Review Letters} \textbf{\bibinfo{volume}{78}},
  \bibinfo{pages}{243} (\bibinfo{year}{1997}).

\bibitem[{\citenamefont{Liboff}(2003)}]{liboff2003ktc}
\bibinfo{author}{\bibfnamefont{R.}~\bibnamefont{Liboff}},
  \emph{\bibinfo{title}{{Kinetic Theory: Classical, Quantum, and Relativistic
  Descriptions}}} (\bibinfo{publisher}{Springer}, \bibinfo{year}{2003}).

\bibitem[{\citenamefont{Helbing and Moln{\'a}r}(1995)}]{helbing1995sfm}
\bibinfo{author}{\bibfnamefont{D.}~\bibnamefont{Helbing}} \bibnamefont{and}
  \bibinfo{author}{\bibfnamefont{P.}~\bibnamefont{Moln{\'a}r}},
  \bibinfo{journal}{Physical Review E} \textbf{\bibinfo{volume}{51}},
  \bibinfo{pages}{4282} (\bibinfo{year}{1995}).

\bibitem[{\citenamefont{Goldhirsch and Ronis}(1983)}]{goldhirsch1983ttg}
\bibinfo{author}{\bibfnamefont{I.}~\bibnamefont{Goldhirsch}} \bibnamefont{and}
  \bibinfo{author}{\bibfnamefont{D.}~\bibnamefont{Ronis}},
  \bibinfo{journal}{Physical Review A} \textbf{\bibinfo{volume}{27}},
  \bibinfo{pages}{1616} (\bibinfo{year}{1983}).

\bibitem[{\citenamefont{Grmela and Garcia-Colin}(1980)}]{grmela1980cek}
\bibinfo{author}{\bibfnamefont{M.}~\bibnamefont{Grmela}} \bibnamefont{and}
  \bibinfo{author}{\bibfnamefont{L.}~\bibnamefont{Garcia-Colin}},
  \bibinfo{journal}{Physical Review A} \textbf{\bibinfo{volume}{22}},
  \bibinfo{pages}{1295} (\bibinfo{year}{1980}).

\bibitem[{\citenamefont{Ottinger}(2005)}]{ottinger2005bet}
\bibinfo{author}{\bibfnamefont{H.~C.} \bibnamefont{Ottinger}},
  \emph{\bibinfo{title}{{Beyond equilibrium thermodynamics}}}
  (\bibinfo{publisher}{Wiley-Interscience, Hoboken, NJ}, \bibinfo{year}{2005}).

\end{thebibliography}

\end{document}